\documentclass[sigconf]{acmart}

\AtBeginDocument{%
  \providecommand\BibTeX{{%
    \normalfont B\kern-0.5em{\scshape i\kern-0.25em b}\kern-0.8em\TeX}}}

\copyrightyear{2022} 
\acmYear{2022} 
\setcopyright{acmlicensed}\acmConference[ICSE '22]{44th International Conference on Software Engineering}{May 21--29, 2022}{Pittsburgh, PA, USA}
\acmBooktitle{44th International Conference on Software Engineering (ICSE '22), May 21--29, 2022, Pittsburgh, PA, USA}
\acmPrice{15.00}
\acmDOI{10.1145/3510003.3510083}
\acmISBN{978-1-4503-9221-1/22/05}

\usepackage{booktabs}
\usepackage{subcaption}
\usepackage[normalem]{ulem}
\usepackage{amsmath}
\usepackage{pifont}
\usepackage{xfrac}
\usepackage[ruled,vlined,linesnumbered]{algorithm2e}
\usepackage{algpseudocode}
\usepackage{url}
\usepackage{comment}
\usepackage{color}
\usepackage{fancyvrb}
\usepackage{balance}
\usepackage{listings}
\usepackage{multirow}
\usepackage{enumitem}
\usepackage{mathtools}
\usepackage{graphicx}
\usepackage{pbox}
\usepackage{makecell}
\usepackage{rotating}
\usepackage{adjustbox}
\usepackage{amsthm}
\usepackage{mathtools}
\usepackage{hyperref}
\usepackage[medium]{titlesec}
\usepackage{bm}
\usepackage{subcaption}

\newcommand{\tool}[0]{\mbox{\textsc{OJXPerf}}}

\begin{document}

\title{\tool{}: Featherlight Object Replica Detection for Java Programs}

\author{Bolun Li}
\affiliation{%
  \institution{North Carolina State University}
  \city{Raleigh}
  \state{North Carolina}
  \country{USA}}
\email{bli35@ncsu.edu}

\author{Hao Xu}
\affiliation{%
  \institution{College of William and Mary}
  \city{Williamsburg}
  \state{Virginia}
  \country{USA}}
\email{hxu07@email.wm.edu}

\author{Qidong Zhao}
\affiliation{%
  \institution{North Carolina State University}
  \city{Raleigh}
  \state{North Carolina}
  \country{USA}}
\email{qzhao24@ncsu.edu}

\author{Pengfei Su}
\affiliation{%
  \institution{University of California, Merced}
  \city{Merced}
  \state{California}
  \country{USA}}
\email{psu9@ucmerced.edu}

\author{Milind Chabbi}
\affiliation{%
  \institution{Scalable Machines Research}}
\email{milind@scalablemachines.org}

\author{Shuyin Jiao}
\affiliation{%
  \institution{North Carolina State University}
  \city{Raleigh}
  \state{North Carolina}
  \country{USA}}
\email{sjiao2@ncsu.edu}

\author{Xu Liu}
\affiliation{%
  \institution{North Carolina State University}
  \city{Raleigh}
  \state{North Carolina}
  \country{USA}}
\email{xliu88@ncsu.edu}

\begin{abstract}
Memory bloat is an important source of inefficiency in complex production software, especially in software written in managed languages such as Java.
Prior approaches to this problem have focused on identifying objects that outlive their life span.
Few studies have, however, looked into whether and to what extent myriad objects of the same type are identical.
A quantitative assessment of identical objects with code-level attribution can assist developers in refactoring code to eliminate object bloat, and favor \emph{reuse} of existing object(s).
The result is reduced memory pressure, reduced allocation and garbage collection, enhanced data locality, and reduced re-computation, all of which result in superior performance.

\sloppy
We develop \tool{}, a \emph{lightweight} sampling-based profiler, which probabilistically identifies identical objects. 
\tool{} employs hardware performance monitoring units (PMU) in conjunction with hardware debug registers to sample and compare field values of different objects of the same type allocated at the same calling context but potentially accessed at different program points.
The result is a lightweight measurement --- a combination of object allocation contexts and usage contexts ordered by duplication frequency.
This class of duplicated objects is relatively easier to optimize.
\tool{} incurs 9\% runtime and 6\% memory overheads on average. 
We empirically show the benefit of \tool{} by using its profiles to instruct us to optimize a number of Java programs, including well-known benchmarks and real-world applications.
The results show a noticeable reduction in memory usage (up to 11\%)  and a significant speedup (up to 25\%).

\end{abstract}

\maketitle

\section{Introduction}
\label{introduction}

Memory bloat is a common problem in managed languages, such as Java and C\#.
The problem is particularly severe in large, production software, which employs layers of abstractions, third-party libraries, and evolves over time into complex systems not comprehendible by any single developer.
Furthermore, these programs run for a long time (several months at a time), giving them an opportunity to grow their memory footprint and become a source of major problems in production environments often shared by several other programs.

An object that is not reclaimed by the garbage collector (GC) but neither read nor written any more is considered to be ``leaked''.
A memory leak happens in managed languages because useless objects remain unreclaimed by the GC because of unnecessary references to them.
Additionally, memory spikes occur in managed languages because of accumulated objects that are yet to be garbage collected.
Memory bloat (whether due to leaks or GC) results in high memory pressure and poor performance.
A lot of prior work exits to detect object leaks~\cite{10.1145/1542476.1542523,10.1145/1806596.1806616,10.1145/1993498.1993530,10.1007/978-3-642-31057-7_32,10.1145/2491411.2491416,10.1145/2509136.2509512,10.1145/2560047,DBLP:conf/cgo/YanXYR14,fang_et_al:LIPIcs:2015:5227} and improve GC~\cite{10.1145/3395659,10.1145/996893.996873,10.1145/1133255.1134021,10.1145/301589.286865,10.1145/1035292.1028983,10.1145/583854.582422}.

However, there is another cause of memory bloat and inefficiency that has hardly been studied --- replica objects --- which is the focus of this paper.
Two objects are replicas if their contents are identical. 
Two objects are \emph{shallow replicas} if their fields are bitwise identical; and 
\emph{deep replicas} if the transitive closure of the constituent objects and their respective fields values are bitwise identical.
When two objects are bitwise identical (shallow replicas), their transitive closures are also the same. However, when two objects are not bitwise identical and the difference occurs on one or more fields that are object references, it becomes necessary to chase those references to disprove that the contents of those objects are not identical. After chasing all object references, if we can prove that their contents are identical, then such two objects are deep replicas.

There is a temporal aspect to replica objects: one may regard two objects as replicas either because they were \emph{identical at the time (or a window) of observation} or for their \emph{entire lifetime}.
Two objects are \emph{mutable replicas} if they are identical to each other, but they may independently evolve during their lifetimes; whereas 
two objects are \emph{immutable replicas} if they are identical to each other and are also immutable during their lifetimes.

There are two performance dimensions to replica objects: total \emph{memory consumption} and total number of \emph{memory accesses}.
On the one hand, an object may be large in size and replicated only a few times, and such replicas still contribute to the overall memory bloat;  also, an object may be small in size but replicated many times, which also contributes to memory bloat; both these cases are worth optimizing to remove the replicas.
On the other hand, an object may be small in size and replicated only a few times, but the program may be accessing these few replicated objects a lot of times; although this situation is not memory bloat, it can still have a significant consequence to the overall performance since it increases the memory footprint at the CPU cache level, squanders potential memory \emph{reuse}~\cite{DBLP:books/mk/AllenK2001}, and often results in redundant re-computations~\cite{10.1145/2491411.2491416, 10.1145/3037697.3037729}.


\begin{figure*}[t]
\centering
\includegraphics[width=0.8\textwidth]{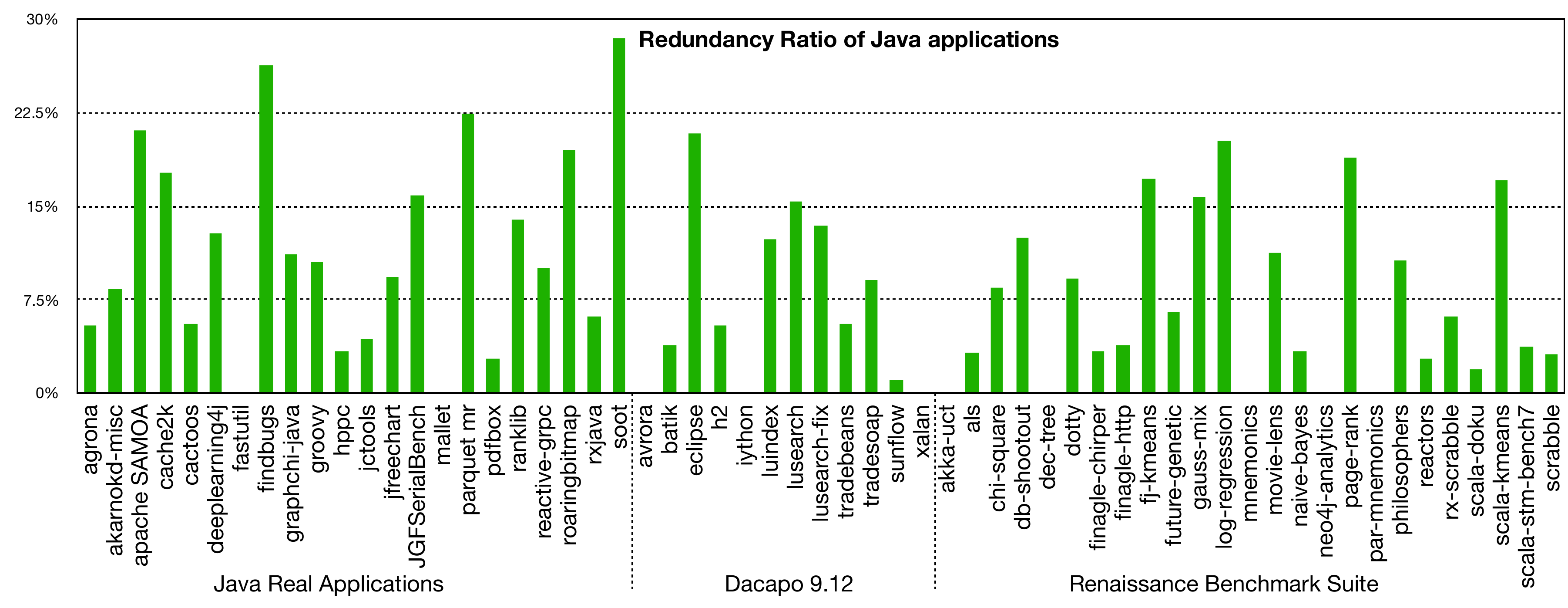}
\caption{Percentage of replicated objects over all the objects in various Java applications.}
\label{redundancy ratio}
\end{figure*}

\begin{figure}
\begin{lstlisting}[firstnumber=90,language=java]
private void readNext() {
  ...
  currentBuffer = new int[currentCount];
  @$\blacktriangleright$@byte[] bytes = new byte[numGroups * bitWidth];
  @$\blacktriangleright$@new DataInputStream(in).readFully(bytes);
  for (int valueIndex = 0, byteIndex = 0; ...; ...) {
    @$\blacktriangleright$@packer.unpack8Values(bytes, byteIndex, currentBuffer, valueIndex);
  }
}
\end{lstlisting}
\vspace{-0.2in}
\captionof{lstlisting}{Object replicas in Parquet MR. The replica object {\tt bytes} is allocated on line 93, initialized on line 94 and used on line 96.}
\label{motivation parquet}
\end{figure}

Having described the landscape of object replication (deep vs. shallow, some time window  vs. full lifespan, mutable vs. immutable, and memory size vs. access counts), we now scope this problem to a tractable subset driven by pragmatic tool-development factors.
First, instrumenting every allocation and memory access to identify object replicas leads to excessive runtime overheads; we seek for a lightweight tool that can collect  profiles in production rather than in test-only environments; we guarantee the analysis accuracy with the theoretical bounds of a sampling technique we use.
Second, deep replica comparison is unachievable without running something analogous to the garbage collector, which can introduce high overheads and require runtime modifications, making it less adaptable; hence we restrict our tool to only shallow object comparison. 
Third, if two objects are not replicas for their entire life span, they are not easy to optimize, and hence we consider only those objects that are replicas for their entire duration. 
We do not enforce objects to be immutable for replica detection.
Finally, our tool regards two or more objects as replicas only if they were allocated in the same calling context; the observation drives this restriction that it is significantly easy to refactor such code to optimize compared with optimizing replica objects allocated from myriad code locations.
We emphasize that we want to be able to monitor replicas and prioritize them by their access frequency.

\tool{}, developed to meet these factors, monitors object allocations and accesses at runtime via statistical sampling.
The key differentiating aspect of \tool{}, compared to a large class of existing profilers, is its ability to detect object replicas with minimal byte code instrumentation and no prior knowledge of programs makes it applicable in the production environment.
A thorough evaluation of several real-world applications shows that pinpointing object replicas offers new avenues to understanding performance losses; aggregating replica objects into one or a few objects reduces the memory footprint, eliminates redundant computations, and enhances performance.


\subsection{Observation}

With the help of \tool{}, Figure~\ref{redundancy ratio} quantifies the ratio of object replicas over the total number of objects in several real applications listed at~\cite{awesome-java} and two popular Java benchmark suites---Dacapo 9.12~\cite{dacapo-9.12} and Renaissance~\cite{Prokopec:2019:RBS:3314221.3314637}, showing that replicated objects are pervasive in modern Java software packages.
Based on many case studies investigated in this paper, we observe that object replication is the symptom of the following kinds of inefficiencies.

\paragraph{\textbf{\textit{Input-sensitive Inefficiencies.}}}
Repeatedly using the same input to instantiate a Java class shows up as repeatedly creating objects with the same content.
Listing~\ref{motivation parquet} shows a problematic method \texttt{readNext} from Parquet-MR~\cite{parquet-mr}, which contains the Java implementation of the Parquet format.
This method is invoked in a loop, and in each invocation, it creates a new object \texttt{bytes}, shown on line 93, and initializes this object via input stream ``in'', as shown on line 94.
We run Parquet-MR using Parquet-Column as its input.
The Parquet-Column input is a columnar storage format for Hadoop; this format provides efficient storage and encoding of data. As line 94 in Listing 1 shows, each time the readNext method is invoked, it creates a copy of input contents (variable “in”), and uses this copy to initialize many objects “bytes” (object replication).
None of the existing profilers, such as JXPerf~\cite{inefficiencies-in-java} and LDoctor~\cite{ldoctor}, can identify such object replicas since they are designed only to recognize the redundancies happening at the same memory location.
Instead, objects \texttt{bytes} are allocated in disjoint memory regions.

\paragraph{\textbf{\textit{Algorithmic Inefficiencies.}}}
Suboptimal choices of an algorithm often show up as object duplications.
As a practical example, Findbugs~\cite{FindBugs} divides a graph into tiny-size blocks and creates an object for each block instead of creating a single object for the whole graph. Consequently, most of the created objects have the same content due to good value locality among adjacent blocks.

\paragraph{\textbf{\textit{Data Structural Inefficiencies.}}}
Like suboptimal algorithms, poor data structures can easily introduce object replicas as well. For instance, in matrix multiplication, sparse matrices with a dense format can yield a high proportion of objects with the same content.\\

The major lessons that can be learned from this paper:
\begin{itemize}[leftmargin=*]
\item Object replicas are not uncommon in real Java applications.
\item Sampling-based measurement based on hardware counters and debug registers can provide good insights and incur significant low overhead. 
\item Developing OJXPerf that efficiently interacts with off-the-shelf JVM and Linux OS in the production environment requires careful design.
\item The call path of object allocation and source code attribution in a GUI are particularly useful for users to identify actionable optimization.
\end{itemize}

\subsection{Paper Contributions}
In this paper, \tool{} makes the following contributions:
\begin{itemize}[leftmargin=*]
\item Develops a novel object-centric profiling technique. It provides rich information to guide optimizing object replicas in JVM-based programs, such as Java and Scala.
\item Employs PMU in conjunction with hardware debug registers and minimal byte code instrumentation, which typically incurs 9\% runtime and 6\% memory overheads.
\item Quantifies the theoretical lower and upper bounds of replication ratios of \tool{}'s statistical approach.
\item Applies to unmodified Java (and languages based on JVM, e.g., Scala) applications, the off-the-shelf Java virtual machine, and Linux operating system, running on commodity CPU processors, which can be directly deployed in the production environment.
\sloppy
\item Provides intuitive optimization guidance for developers. We evaluate \tool{} with popular Java benchmarks (Dacapo~\cite{dacapo}, NPB~\cite{npb}, Grande~\cite{grande}, SPECjvm2008~\cite{specjvm2008}, and the most recent Renaissance~\cite{Prokopec:2019:RBS:3314221.3314637}) and more than 20 real-world applications. Guided by \tool{}, we are able to obtain significant speedups by eliminating object replicas in various Java programs. We have upstreamed some of the patches to the software repositories.
\end{itemize}

\subsection{Paper Organization}
The paper is organized as follows. Section~\ref{relate} covers the related work and distinguishes \tool{}. Section~\ref{background} offers some background knowledge. Section~\ref{methodology} depicts \tool{}'s methodology. Section~\ref{implementation} describes the implementation details of \tool{}. Section~\ref{theory} discusses the theoretical guarantee of \tool{}'s analysis accuracy. Section~\ref{evaluation} evaluates \tool{}'s accuracy and overhead. Section~\ref{case study} describes case studies of \tool{}. Section~\ref{validity} discusses the threats to validity. Finally, Section~\ref{conclusion} presents our conclusions.

\section{Related Work}
\label{relate}
Performance profiling techniques abound in the Java community, which fall into two categories: hardware and software approaches. Each category can be further classified into hotspot and inefficiency analyses. We also compare the related Java tools in table~\ref{comparison}. 

\begin{table}
\centering
\includegraphics[width=0.49\textwidth]{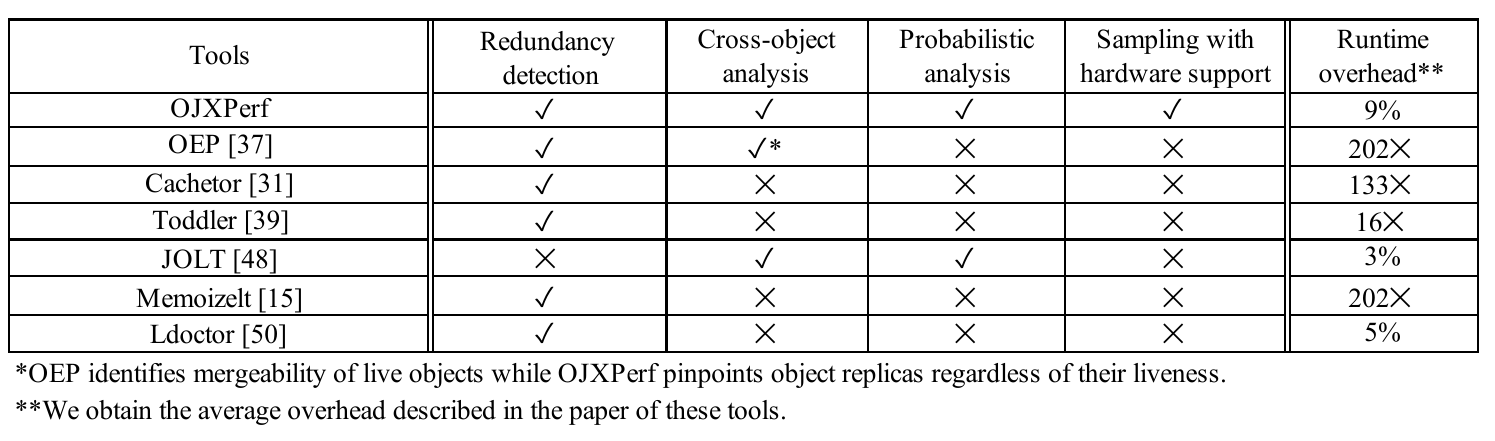}
\caption{Comparing OJXPerf with other state-of-the-art inefficiency analysis tools/approaches.}
\label{comparison}
\vspace{-1em}
\end{table}

\subsection{Software Approaches}
\paragraph{\textbf{\textit{Hotspot Analysis.}}}
\sloppy
Netbeans Profiler~\cite{netbeans-WWW}, JProfiler~\cite{jprofiler-WWW}, YourKit~\cite{yourkit-WWW}, VisualVM~\cite{visualvm-WWW}, and Oracle Developer Studio Performance Analyzer~\cite{oracle-studio-WWW} are hotspot analysis profilers, which identify execution hotspots in CPU time or memory usage. They typically introduce negligible overhead by leveraging OS timers as the sampling engines to deliver periodic samples.
The hotspot analysis is indispensable but fails to tell whether a resource is being used in a productive manner and contributes to a program's overall efficiencies.
A hotspot does not need to be an inefficient code region and vice versa.
Hence, a heavy burden is on users to make a judgment on whether the reported hotspots are actionable.

\paragraph{\textbf{\textit{Inefficiency Analysis.}}}
Unlike hotspot analysis, inefficiency analysis tools identify code regions leading to resource wastage instead of resource usage.
Cachetor~\cite{Cachetor} combines value profiling and dependence profiling to pinpoint operations that repeatedly generate an identical value.
MemoizeIt~\cite{Memoization} identifies methods that repeatedly perform identical computation.
JOLT~\cite{LJOLT} identifies and optimizes object churn in a virtual machine.
Toddler~\cite{toddler} identifies redundant memory load operations in loop nests.
The follow-up work~\cite{ldoctor} applies a static-dynamic analysis to reduce Toddler's overhead. However, it identifies inefficiencies within a small number of suspicious loops instead of the entire program.
Xu \textsl{et al.}~\cite{Flow} introduce copy profiling that optimizes data copies to remove the objects that carry copied values, and the method calls that allocate and populate these objects. Their follow-up work~\cite{Containers} develops practical static and dynamic analyses that identify inefficiently-used containers, such as overpopulated containers and underutilized containers. They also present a run-time technique~\cite{Reusable} to identify reusable data structures to avoid frequent object allocations. OEP~\cite{OEP} identifies mergeability among live objects, which requires the measurement of object reachability. In contrast, OJXPerf analyzes objects allocated in the same call path regardless of their liveness. OEP leverages bytecode instrumentation, which incurs orders of magnitude of overhead compared to OJXPerf.

\tool{} is a profiler but applies a hardware approach to address a different inefficiency problem --- object replication. 

\subsection{Hardware Approaches}
There are many hardware-assisted profilers. In this paper, we review only PMU- or debug register-assisted Java profilers.

\paragraph{\textbf{\textit{Hotspot Analysis.}}}
Linux Perf~\cite{perf}, Async-profiler~\cite{async-profiler-WWW}, and Oprofile~\cite{Levon:OProfile} employ PMU as the sampling engines to deliver periodic samples. PMU-based hotspot profilers offer slightly better intuition than the OS timer-based ones since they can classify hotspots according to various forms of performance metrics collected from PMU, such as instruction numbers, cache misses, bandwidth, and many others. However, they are not panaceas; users still have to distinguish inefficient hotspots from efficient ones manually.

\paragraph{\textbf{\textit{Inefficiency Analysis.}}}
Sweeney \textsl{et al.}~\cite{pmu-java-behavior} develop a system that provides a graphical interface to alleviate the difficulty in interpreting PMU results. Hauswirth \textsl{et al.}~\cite{Vertical} present vertical profiling that captures and correlates performance problems across multiple execution layers (application, VM, OS, and hardware). Georges \textsl{et al.}~\cite{Phase} study methods exhibit similar and dissimilar behaviors by measuring the execution time for each method invocation using PMU. Lau \textsl{et al.}~\cite{auditing} present a technique that allows a VM to determine whether an optimization improved or degraded by measuring CPU cycles. Remix~\cite{remix} employs PMU to identify inter-thread false sharing on the fly.
JXPerf~\cite{inefficiencies-in-java} detects redundant memory operations by using PMU to sample memory locations accessed by a program and using debug registers to monitor subsequent accesses to the same location.

Orthogonal to the aforementioned inefficiency analysis profilers, \tool{} addresses a different inefficiency problem with a different usage of PMU and debug registers. To the best of our knowledge, \tool{} is the first lightweight sampling-based profiler to pinpoint object replicas in Java.

\section{Background}
\label{background}

We introduce some essential facilities that \tool{} leverages based on Java virtual machines (JVM) and CPU processors.

\paragraph{\textbf{\textit{Java Virtual Machine Tool Interface (JVMTI).}}}
JVMTI, a native programming interface of the JVM, is loaded during the initialization of the JVM. 
JVMTI provides a VM interface for the full breadth of tools that need access to VM state, including but not limited to profiling, debugging, monitoring, thread analysis, and coverage analysis tools. 

\paragraph{\textbf{\textit{Hardware Performance Monitoring Unit (PMU).}}}
PMU is hardware built inside a processor to measure its performance parameters. We can measure parameters like instruction cycles, cache hits, cache misses, branch misses, and many others, depending on the supported hardware. PMU supports lightweight measurement. 

Intel processors also support Precise Event-Based Sampling (PEBS)~\cite{IntelArch:PEBS:Sept09}. PEBS is a profiling mechanism that logs a snapshot of the processor state at the time of the event, allowing users to attribute performance events to actual instruction pointers (IPs). Also, PEBS provides an effective address (EA) at the time of the sample when the sample is for a memory load or store instruction. 
In PEBS, the event type may be chosen from an extensive list of performance-related events to monitor, e.g., cache misses, remote cache hits, branch mispredictions. AMD processors provide similar capabilities via instruction-based sampling.

\paragraph{\textbf{\textit{Hardware Debug Register.}}}
Modern x86 processors provide debug facilities for developers in debugging code and monitoring system behaviors. Such debug support is accessed using hardware debug registers. Hardware debug registers~\cite{software-debugging, debuggable} enable trapping the CPU execution for debugging when the program counter (PC) reaches an address (breakpoint) or an instruction accesses a designated address (watchpoint). Hardware debug registers allow programmers to selectively enable various debug conditions associated with a set of four debug addresses because our current x86 processors have four debug registers.

\section{Methodology}
\label{methodology}

As  previously alluded, we restrict the definition of object replicas to those allocated in the same calling context.

\begin{definition}\textbf{Object Replicas:}
\textit{$O_1$ and $O_2$ are two objects that have the same allocation context. 
If the contents of $O_1$ and $O_2$ are identical, $O_1$ is a replica of $O_2$ and $\langle O_1, O_2\rangle$ is an object replication pair.} 
\end{definition}

A straightforward detection approach is monitoring every allocation context and comparing all fields of all object instances created at that allocation context at any use points. However, performing such whole-heap object tracing can introduce a prohibitively high overhead ($70$\textasciitilde $300\times$ slowdown reported in~\cite{Merlin}).

Instead of exhaustive duplication detection, \tool{} takes advantage of sampling to perform lightweight replica detection.
We neither compare all objects allocated in the same context nor compare all fields when comparing two objects.
Instead, our algorithm chooses random fields (offsets in terms of memory locations) at random points.

Example 1 shows an object replica detection example that keeps allocating an object $O$ inside a while loop. As the while loop iterates $4$ times, the program allocates a sequence of objects $\{O_1, O_2, O_3, O_4\}$, which have the same allocation context and are sorted by the allocation timestamp during program execution. 
First, in each loop iteration, we intercept the allocation of the object on line 3. 
This interception offers two pieces of information: 1) the calling context of allocation, and 2) the memory address range occupied by each object.
We maintain this information for future use.
For all objects allocated on line 3 in this example, the context is the same but the object addresses can be different\footnote{two objects of different size, e.g., arrays allocated in the same context are easily ruled out of duplication due to size difference}.
Second, when the program accesses an object (use point), e.g., lines 5 and 7, we can obtain the effective address of the access and map the address to the object it belongs to; furthermore, we can easily derive the relative offset of the access from the start address of the object; this relative offset guides us where to monitor another object allocated in the same context.
Third, when the program accesses an object, we can read the contents of the location accessed.

\renewcommand{\thealgocf}{}
\renewcommand{\algorithmcfname}{Example 1}
\begin{algorithm}[h]
\caption{Example of Object Replica Detection.}
\SetAlgoLined
i = 0\;
\While{i < 4}{
  allocate an object O\;
  initialize O\;
  use O; //use O for the first time\\
  update O\;
  use O; //use O for the second time\\
  i++\;
}
\end{algorithm}

\begin{figure}[t]
\centering
\includegraphics[width=0.49\textwidth]{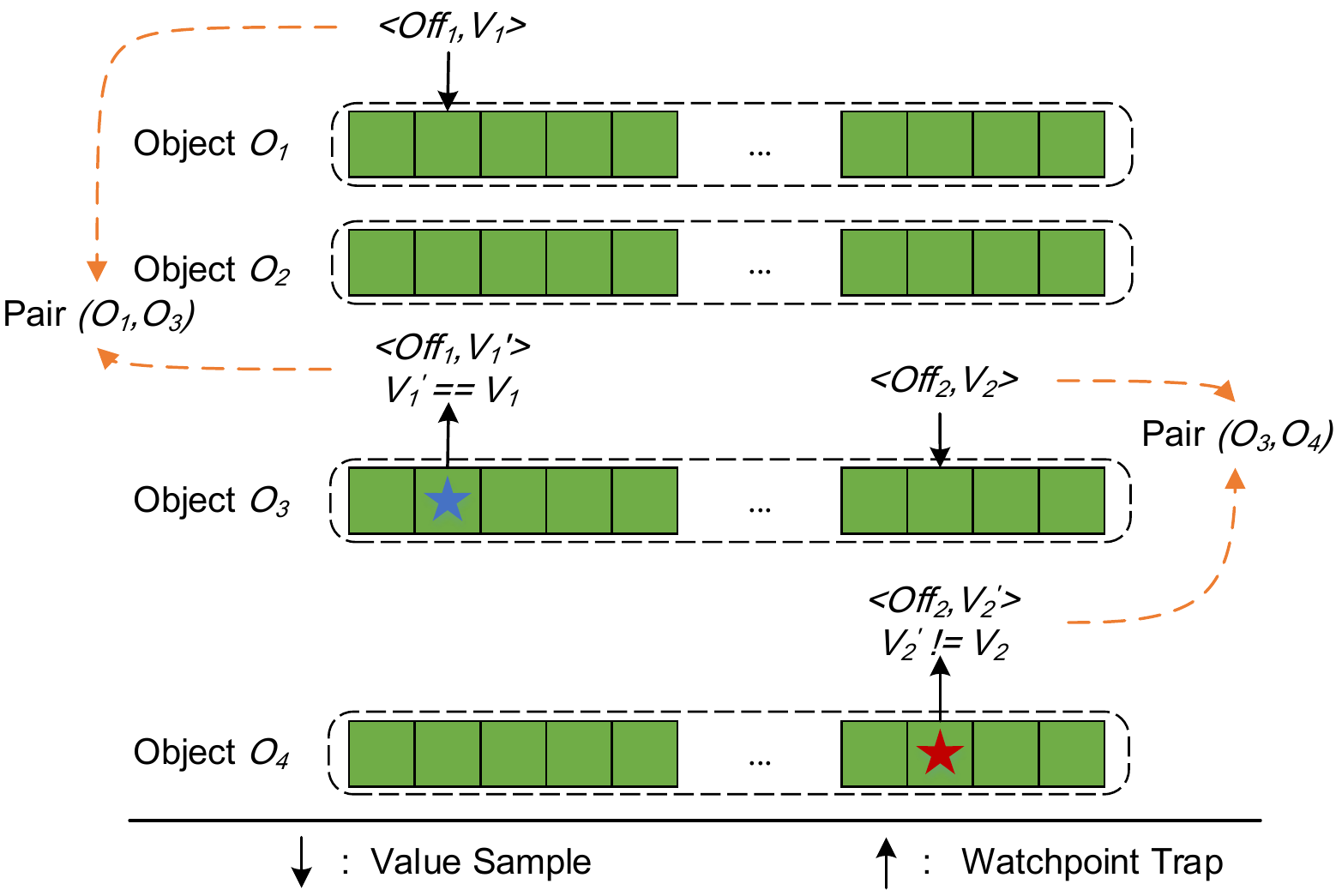}
\caption{Watchpoint scheme for object replica detection. $\mathit{Off_1}$ ($\mathit{Off_2}$) presents memory offset with value $V_1$ ($V_2$) for Object $O_1$ (Object $O_3$). When a watchpoint trap of memory access happens at offset $\mathit{Off_1}$ ($\mathit{Off_2}$), \tool{} compares their corresponding values $V_1$ and $V_1'$ ($V_2$ and $V_2'$).}
\label{watchpoint}
\vspace{-1em}
\end{figure}

Given the nature of sampling, let's assume that memory-access samples occur on line 5 in iteration 1 of the loop (while accessing object $O_1$) and line 7 in iteration 3 (while accessing object $O_3$), as shown in Figure~\ref{watchpoint}.
Assume, the sample in iteration 1 for $O_1$ on line 5 happens at the relative offset  $\mathit{Off_1}$ from the beginning of the object and the value at $\mathit{Off_1}$ is $V_1$; \tool{} remembers the triple --- line 5, $\mathit{Off_1}$, and $V_1$ --- for future use.
For the next allocation, $O_2$, we probabilistically skip and do nothing.
When $O_3$ starts to be accessed, we decide to monitor its contents at offset $\mathit{Off_1}$, and hence arm a watchpoint to trap on access to offset $\mathit{Off_1}$ from the beginning of $O_3$.
This watchpoint traps when the program accesses $O_3$ on line 5.
Let the contents at $O_3 + \mathit{Off_1}$ be $V_1'$ when the trap happens.
We compare $V_1$ and $V_1'$ and if they are the same,  $O_1$ and $O_3$ contribute towards the number of equivalent objects allocated in context \texttt{line 3};
otherwise they contribute towards non-equivalent objects allocated in context \texttt{line 3}.

The next sample happens on line 7, for the same object $O_3$ at offset $\mathit{Off_2}$ in iteration 3 of the loop.
Let the value at $\mathit{Off_2}$ for $O_3$ be $V_2$.
We remember the triple --- line 7, $\mathit{Off_2}$, and $V_2$ --- for future use.
When $O_4$ starts to be accessed in iteration 4 of the loop, we arm a watchpoint at address $\mathit{Off_2}$ from the beginning of $O_4$.
This watchpoint traps when the program accesses $O_4$ on line 7.
Let the value at the trapped location be $V_2'$.
As before, depending on  whether $V_2$ and $V_2'$ are the same or not, they contribute towards equivalent or non-equivalent objects allocated in context \texttt{line 3}. 
Figure~\ref{watchpoint} shows that $V_1$ equals to $V_1'$ (the blue star), which means that the values stored in an offset $\mathit{Off_1}$ of $O_1$ and $O_3$ are the same. Also, the red star in Figure~\ref{watchpoint} shows that the values stored in an offset $\mathit{Off_2}$ of $O_3$ and $O_4$ are different ($V_2$ doesn't equal to $V_2'$), which means that $O_3$ and $O_4$ must be two objects that have different contents.

As the program continues, \tool{} performs the same redundancy checks for other samples taken from objects $\{O_1, O_2, O_3, O_4\}$.
Finally, if most of the comparisons (>60\%, obtained from our experiments) report identical values among all detection pairs, we believe objects $\{O_1, O_2, O_3, O_4\}$ suffer from object replicas with a high probability, which is quantified with our theoretical analysis in Section~\ref{theory}.

\section{Implementation}
\label{implementation}

\begin{figure}[t]
\centering
\includegraphics[width=0.49\textwidth]{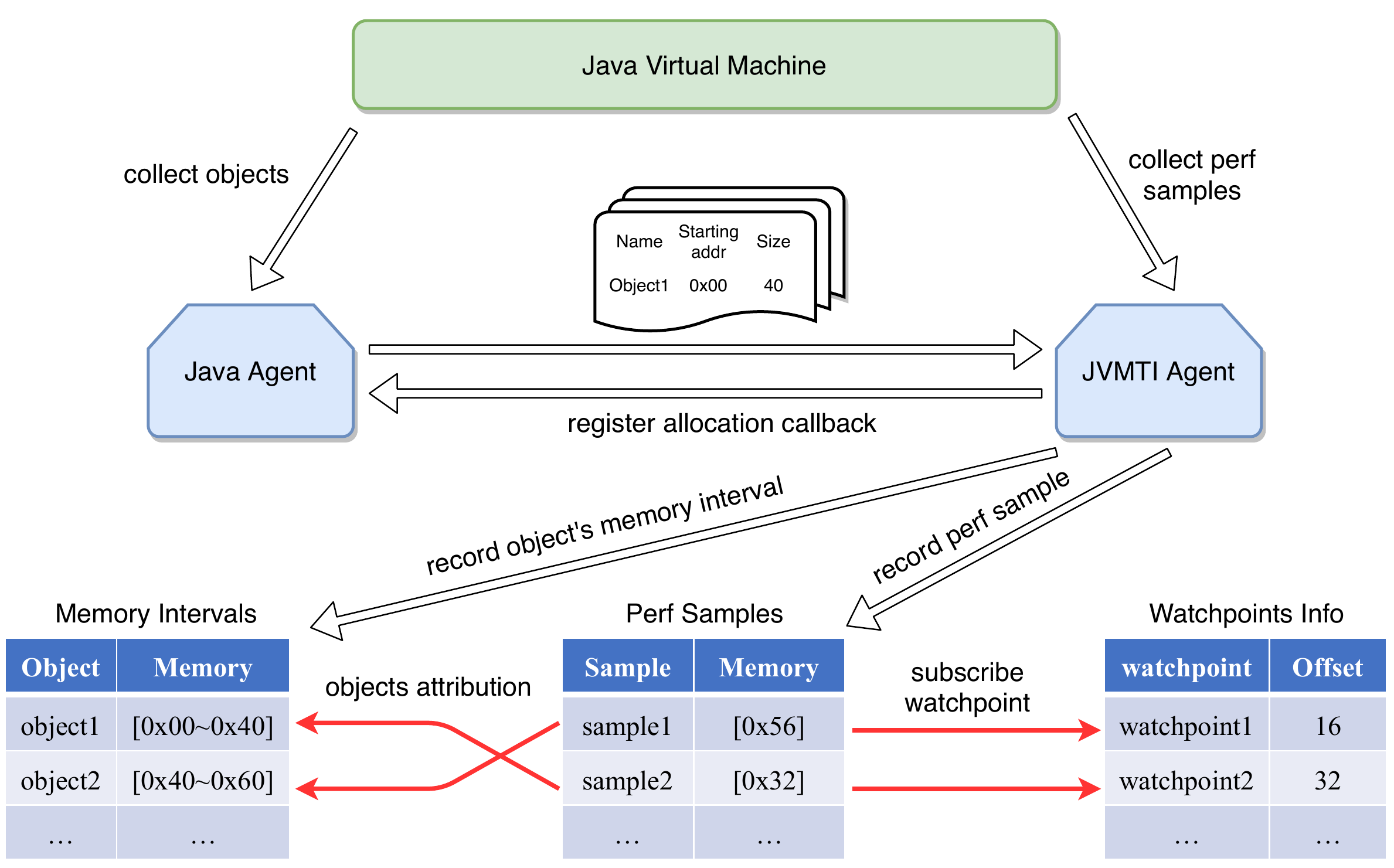}
\caption{Overview of \tool{}'s profiling.}
\label{communication}
\end{figure}

\tool{} is a user-space tool with no need for any privileged system permission.
\tool{} requires no modification to hardware, OS, JVM, and monitoring applications, making it applicable to the production environment. 
Conceptually, \tool{} consists of two components: data-centric analysis and duplication detection. These two components are implemented within two agents: a Java agent and a JVMTI agent. Figure~\ref{communication} overviews the design of these two agents. The Java agent instruments Java byte code execution to obtain each object's memory interval and allocation context. The JVMTI agent subscribes to Java thread creation to enable PMU. Upon each PMU sample, \tool{} obtains the effective address of the monitored memory access and associates it with the Java object enclosing this address. Moreover, to identify object replicas, the JVMTI agent programs the debug registers to subscribe to watchpoints. 

\subsection{Java Agent}
\label{subsec:data-centric}
The Java agent monitors object allocation, which leverages {\tt java.lang.instrument} API and {\tt ASM} framework. The Java agent inserts pre- and post-allocation hooks to intercept each object allocation. Then, a user-defined callback is invoked on each allocation to obtain the object information, such as the object pointer, type, and size. For a given Java class we want to instrument, the Java agent scans the byte code of this class, instruments {\tt new}, {\tt newarray}, {\tt anewarray}, and {\tt multianewarray}, and obtains the memory range of every object following an existing technique~\cite{object-addr}.

How to present an object allocation is a challenging question.
We adopt a simple and perhaps the most intuitive approach that developers can identify with --- the allocation context leading to the object allocation.
A Java application can often create multiple object instances via a single allocation site in a loop.
All those objects will be represented by a single call path, which naturally aggregates numerous objects with similar behavior. \tool{} leverages the \texttt{AsyncGetCallTrace()}~\cite{AsyncGetCallTrace-WWW} API provided by Oracle Hotspot JVM to determine the calling contexts at any point during the execution.
\tool{} then inserts a $\langle key, value\rangle$ pair into a map $\mathcal{M}$, where $key$ is the memory range and $value$ is the allocation context.

\subsection{JVMTI Agent}
\paragraph{\textbf{\textit{Implementing Sampling with PMU.}}}
The JVMTI agent leverages PMU to sample memory accesses. It subscribes to {\tt MEM\_UOPS\_\\RETIRED:ALL\_LOAD}, a PMU precise event to sample memory loads. 
We empirically choose a sampling period to ensure \tool{} can collect 20-200 samples per second per thread, which yields a fair tradeoff between runtime overhead and statistical accuracy~\cite{Tallent:2010:phd}.
Moreover, the JVMTI agent captures the calling contexts for both PMU samples and object allocations described in Section~\ref{subsec:data-centric}. 
To minimize synchronization, each thread collects PMU samples independently and maintains a thread-local compact calling context tree (CCT)~\cite{Arnold-Sweeney:1999:cct}, which stores the calling contexts of PMU samples and merges all the common prefixes of given calling contexts.

\begin{figure}
  \begin{minipage}[b]{0.49\textwidth}
    \includegraphics[width=1\textwidth]{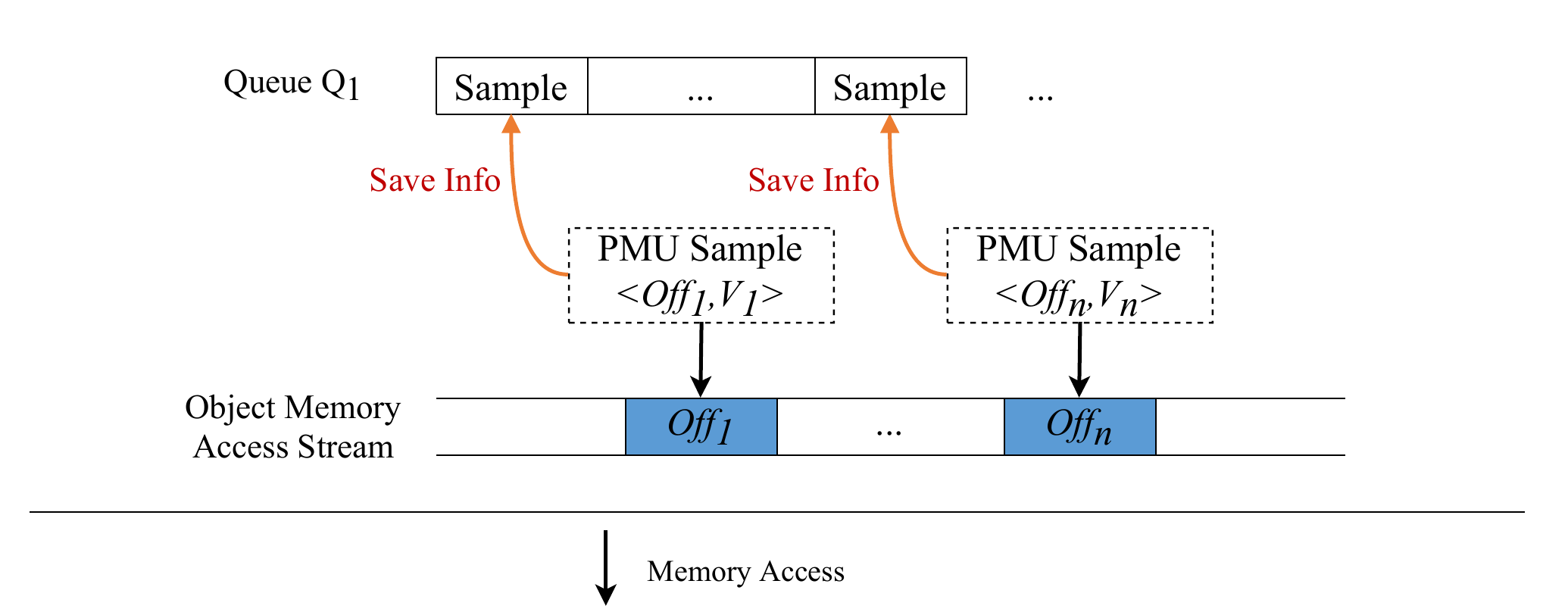}
    \subcaption{The workflow of object replica detection: collecting PMU samples taken from object $O_1$.}
    \label{workflow_a}
  \end{minipage}
 
  \begin{minipage}[b]{0.49\textwidth}
    \includegraphics[width=1\textwidth]{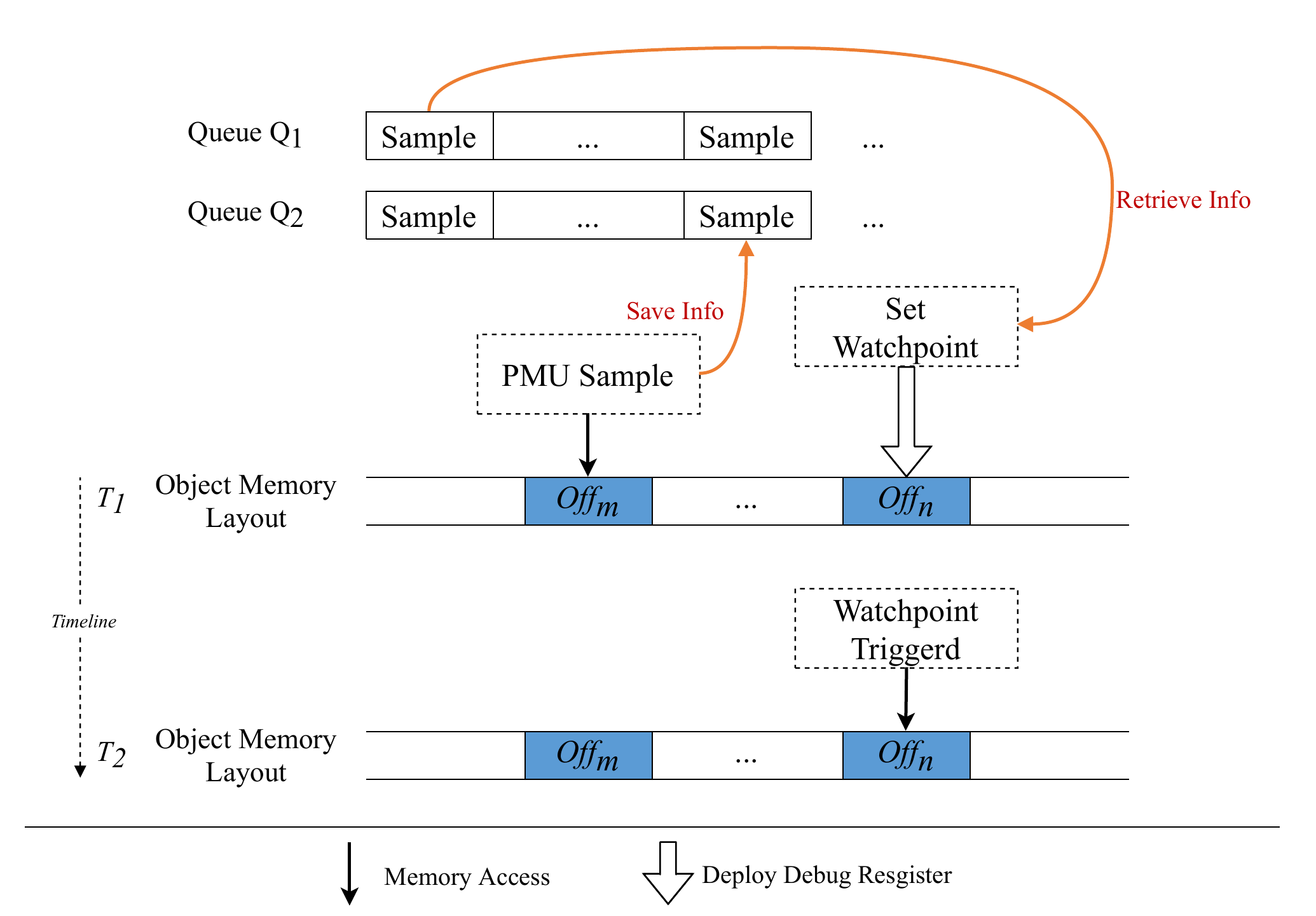}
    \subcaption{The workflow of object replica detection: setting up watchpoint at object $O_2$.}
    \label{workflow_b}
  \end{minipage}
  \caption{Workflow of object-level redundancy detection.}
\end{figure}

\paragraph{\textbf{\textit{Examining Object Contents with Watchpoints.}}}
\tool{} leverages debug registers to set up watchpoints, which traps the program execution when the designated memory addresses are accessed. 
Assume $O_1$ and $O_2$ are two distinct objects that have the same allocation context and $O_1$ is created prior to $O_2$. 
Moreover, $O_1$ and $O_2$ have the same accessing context, then $O_1$ and $O_2$ form a $pair(O_1, O_2)$ as object replicas.
\tool{} uses queues $Q_1$ and $Q_2$ to store samples taken from objects $O_1$ and $O_2$, respectively.
Upon a sample taken from $O_1$, \tool{} uses a tuple $\langle \mathit{Off}, V\rangle$ to represent it and adds this tuple to queue $Q_1$, as shown in Figure~\ref{workflow_a}. 
$\mathit{Off}$ is the offset between the sampled address (i.e., the address of the PMU sample) and the starting address of $O_1$, and $V$ is the value stored at the sampled memory address.
Upon a sample taken from $O_2$, \tool{} not only adds a tuple $\langle \mathit{Off_m}, V_m\rangle$ to queue $Q_2$, but also retrieves a sample ($\langle \mathit{Off_n}, V_n\rangle$) from queue $Q_1$ and uses a debug register to set up a watchpoint at the offset $\mathit{Off_n}$ of $O_2$, as shown in Figure~\ref{workflow_b}. 
\tool{} compares values at the the same offset $\mathit{Off_n}$ of $O_1$ and $O_2$ when the watchpoint is triggered. Watchpoint can be removed when it is triggered, and watchpoints are used for a single access.

\paragraph{\textbf{\textit{Limited Number of Debug Registers.}}}

Hardware offers only a small number of debug registers, which becomes a limitation if the PMU delivers a new sample, but all watchpoints are armed with addresses obtained from prior samples. \tool{} employs a reservoir sampling strategy~\cite{reservoir}, which uniformly chooses between old and new samples with no bias. The basic idea of reservoir sampling is to assign a probability to each debug register and perform a replacement policy based on the probability. Prior work~\cite{featherlight, witch} has shown that reservoir sampling guarantees the fairness of the measurement with a limited number of debug registers. 

\subsection{Offline Data Analyzer and GUI}
To generate a compact profile, which is essential for analyzing a large-scale execution, the offline data analyzer merges profiles from different threads. 
Object allocation call paths coalesce across threads in a top-down way if they are identical.
All memory accesses with their call paths to the same objects are merged as well. 
Metrics are also summed up when call paths coalesce. 
The offline procedure typically takes less than one minute in our experiments.
Furthermore, \tool{} integrates its analysis visualization in Microsoft Visual Studio Code, which is shown in Figure~\ref{gui}.

\subsection{Discussions}

As a sampling-based approach, \tool{} may introduce false positives and false negatives, which are elaborated in Section~\ref{false positive}. Our theoretical analysis to be described in the next section bounds the analysis accuracy.

\section{Theoretical Analysis}
\label{theory}
Since \tool{} does not exhaustively check every field of an object for replication due to the sampling, we compute the lower and upper bounds of the analysis to quantify the replication factor.
\begin{definition}\textbf{Replication Factor (RF) \bm{$\theta$}:}
\textit{For a set of objects that suffer from object replication, the replication factor $\theta$ is the probability of last accessed object to be bit-wise same as the current accessed object. We define $\theta$ as the ratio of the number of times that an object accessed is equivalent to another object accessed previously, to the total accesses of this set of objects.}
\begin{eqnarray}
\scriptsize
\begin{aligned}
\theta=&{\text{num equivalent access times}\langle Object \mathcal{O}\rangle \over {\text{num equivalent + num different access times} \langle Object \mathcal{O}\rangle}}
\end{aligned}
\end{eqnarray}
\end{definition}

Assume $O_1$ and $O_2$ are a pair of object under object replica detection. If all memory locations sampled from $O_2$ have the same values as the corresponding locations in $O_1$ ($O_2^{offset} = O_1^{offset}$), it is possible that $O_2$ and $O_1$ are two objects that have the same contents ($O_2 \equiv O_1$) or the different contents ($O_2 \nequiv O_1$). Here, we have three scenarios and each with a specific probability:

\begin{itemize}
	\item $O_2^{offset} = O_1^{offset}$ and $O_2 \equiv O_1$, the probability is $A$;
	\item $O_2^{offset} = O_1^{offset}$ and $O_2 \nequiv O_1$, the probability is $B$;	
	\item $O_2^{offset} \ne O_1^{offset}$, so $O_2 \nequiv O_1$, the probability is $C$.
\end{itemize}
Obviously, we have $A + B + C = 1$.

Then, the $\theta$ can be rewritten using $A, B, C$ as:
\begin{equation}
\theta = \frac{A + B}{A + B + C} = A + B
\label{eqn:theta}
\end{equation}

Furthermore, we define $\alpha$ as the probability of $O_2^{offset} = O_1^{offset}$ when $O_2 \nequiv O_1$. Then $\alpha$ can be denoted as:

\begin{equation}
\alpha = \frac{ B}{B + C} \ge  \frac{B}{A + B + C} = B
\label{eqn:alpha}
\end{equation}

Combine Equation (\ref{eqn:theta}) and Inequality (\ref{eqn:alpha}), we have:
\begin{equation}
A = \theta - B \ge \theta - \alpha
\label{thetaalpha}
\end{equation}

Assume there are $X$ objects $\{O_1, O_2, ..., O_x\}$ belonging to the same calling context. These $X$ objects are divided into $N$ groups. Inside each group, the objects are identical with each other. Every group contains $X_n$ objects ($1 \le n \le N, \sum_{n=1}^{N}{X_n} = X$), and $X_1, X_2, ..., X_N$ are sorted in an ascending order by group size. 
Based on these $X$ objects, there are ${X \choose 2}$ object pairs. Among these ${X \choose 2}$ object pairs, there will be $\sum_{n= 1}^{N}{X_n \choose 2}$ identical object pairs. Considering we can estimate identical object pairs ratio by $\frac{A}{A+B+C} = A$ and Inequality (\ref{thetaalpha}), we can state:

\begin{equation}
\small
\frac{\sum_{n= 1}^{N}{{X_n \choose 2}}}{{X \choose 2}} = A \ge \theta - \alpha
\label{eqn:fracion}
\end{equation}

Then, $\frac{\sum_{n= 1}^{N}{{X_n \choose 2}}}{{X \choose 2}}$ can be derived  as:
\begin{equation}
\small
\begin{split}
\frac{\sum_{n= 1}^{N}{{X_n \choose 2}}}{{X \choose 2}} = \frac{\sum_{n= 1}^{N}{X_n(X_n - 1)}}{X(X-1)}
& < \frac{\sum_{n= 1}^{N}{X_n^2}}{X^2}
\end{split}
\label{eqn:fracionbound}
\end{equation}
Focusing on $\sum_{n=1}^{N}{(\frac{X_n}{X})^2}$, we can reconstruct it as:
\begin{equation}
\small
\begin{split}
\sum_{n=1}^{N}{(\frac{X_n}{X})^2} &= \sum_{n=1}^{N-1}{(\frac{X_n}{X})^2 + \frac{X_N}{X} * \frac{X_N}{X}} \\
&= \sum_{n=1}^{N-1}{(\frac{X_n}{X})^2 + (1 - \sum_{n=1}^{N-1}{\frac{X_n}{X}}) * \frac{X_N}{X}} \\
&=  \frac{X_N}{X} - \sum_{n=1}^{N-1}{\frac{X_n}{X}(\frac{X_N - X_n}{X} )}
\end{split}
\label{eqn:fracionreconstruct}
\end{equation}
Since $X_N > X_{N-1} >  X_{N-2} > ... > X_1$, we have \begin{equation} \small
\sum_{n=1}^{N-1}{\frac{X_n}{X}(\frac{X_N - X_n}{X} )} > 0
\label{eqn:smallvalue}
\end{equation}
Furthermore, combining (\ref{eqn:fracion}), (\ref{eqn:fracionbound}), (\ref{eqn:fracionreconstruct}), (\ref{eqn:smallvalue}), we then obtain
$$\frac{X_N}{X} \ge \theta - \alpha + \sum_{n=1}^{N-1}{\frac{X_n}{X}(\frac{X_N - X_n}{X} )} > \theta - \alpha,$$
Because $\frac{X_N}{X}$ represents the largest identical objects group size ratio, we know that this ratio is lower bounded by $\theta - \alpha$. Next we show the upper bound of $\frac{X_N}{X}$.

Based on equation \ref{eqn:fracion}, we have:

\begin{equation}
\frac{\sum_{n= 1}^{N}{{X_n \choose 2}}}{{X \choose 2}} = A > \frac{{X_N \choose 2}}{{X \choose 2}}
\label{eqn:fracion_upperbound}
\end{equation}

Focusing on $\frac{{X_N \choose 2}}{{X \choose 2}}$, we have:
\begin{equation}
\begin{split}
\frac{{X_N \choose 2}}{{X \choose 2}} = \frac{X_N(X_N-1)}{X(X-1)} = \frac{X_N}{X} (\frac{X_N}{X} + \frac{X_N-1}{X - 1} - \frac{X_N}{X}) \\
= \frac{X_N}{X} (\frac{X_N}{X} - \frac{X- X_N}{X(X - 1)} )
\end{split}
\label{eqn:fracion_simplification}
\end{equation}

We also have:
\begin{equation}
\frac{X - X_N}{X(X-1)} = \frac{1}{X-1} - \frac{1}{X-1} * \frac{X_N}{X}
\label{eqn:fracion_decompose}
\end{equation}

We then denote $\frac{X_N}{X} = s$ and $\frac{1}{X - 1}  = t$, equation \ref{eqn:fracion_simplification} can be rewritten as:
\begin{equation}
\frac{{X_N \choose 2}}{{X \choose 2}} = s(s-t + st)
\label{eqn:fracion_rewrite}
\end{equation}

Combining with equation \ref{eqn:fracion_upperbound}, we have this in-equation:
\begin{equation}
A > s(s-t + st) = (t + 1)s^2 - st > s^2 - st 
\label{eqn:fracion_inequation}
\end{equation}

Solving this in-equation, we then have:
\begin{equation}
s < \frac{t + \sqrt{t^2 + 4A}}{2}
\label{eqn:solve_inequation}
\end{equation}

Focusing on $A$ in equation \ref{eqn:theta} and \ref{eqn:alpha}, we have:
\begin{equation}
A = \theta - B = \theta - \alpha * (B + C) = \theta - \alpha * (1 - A)
\label{eqn:A}
\end{equation}

Solving it, we then have:
\begin{equation}
A = \frac{\theta -\alpha} {1 - \alpha}
\label{eqn:A_solved}
\end{equation}

Based on the in-equation \ref{eqn:solve_inequation}, we have:
\begin{equation}
\begin{split}
s < \frac{t + \sqrt{t^2 + 4\frac{\theta-\alpha}{1-\alpha}}}{2} = t/2 + \sqrt{t^2/4 + \frac{\theta-\alpha}{1-\alpha} } \\
= \frac{1}{2(X-1)} + \sqrt{\frac{1}{4(X-1)^2} + \frac{\theta-\alpha}{1-\alpha}}
\end{split}
\label{eqn:upper_bound}
\end{equation}

Here, we have both lower bound and upper bound of $\frac{X_N}{X}$:
$\theta - \alpha < \frac{X_N}{X} < \frac{1}{2(X-1)} + \sqrt{\frac{1}{4(X-1)^2} + \frac{\theta-\alpha}{1-\alpha}}$

For real applications, usually we have $X >> 1$, so $\frac{1}{X-1} \rightarrow 0$. And also we have $\frac{\theta - \alpha}{1 - \alpha} < \frac{\theta}{1} = \theta$.

\begin{definition}\textbf{Lower Bound Factor (LBF) \bm{$\omega$}:}
\textit{$\omega$ is defined as the lower bound of the largest identical objects group size ratio $\frac{X_N}{X}$, so we have $\omega = \theta - \alpha$. }
\end{definition}

\begin{definition}\textbf{Upper Bound Factor (UBF) \bm{$\gamma$}:}
\textit{$\gamma$ is defined as the upper bound of the largest identical objects group size ratio $\frac{X_N}{X}$, so we have $\gamma = \frac{1}{2(X-1)} + \sqrt{\frac{1}{4(X-1)^2} + \frac{\theta-\alpha}{1-\alpha}}$. }
\end{definition}

We show how the interval of the largest identical objects group size ratio $\frac{X_N}{X}$ guides our optimizations in Section~\ref{evaluation}.

In the applications we evaluated, we have not seen an application with a very high $\alpha$ (we compute $\alpha$ for each application via exhaustively checking every field of objects), which we further discuss here. For an application with inequivalent objects $X_1$, $X_2$, ..., $X_N$ that belong to the same calling context, high $\alpha$ indicates that most of the contents of these objects are the same and only a few contents are different, which means these objects are partially replicated. It is worth noting that partially replicated objects can warrant some optimization to move redundant computations, such as approximate computing or data compression; however, it is out of the scope of this paper.

The theoretical analysis influences the design decisions in two aspects: on the one hand, the theoretical bounds guarantee the analysis accuracy of \tool{}'s sampling technique; on the other hand, the bounds, as metrics, help users determine whether the object replicas are significant for optimization.

\section{Evaluation}
\label{evaluation}

We evaluate \tool{} on a 36-core Intel Xeon E5-2699 v3 (Haswell) CPU clocked at 2.3GHz running Linux $4.8.0$.
The memory hierarchy consists of a private 32KB L1 cache, a private 256KB L2 cache, a shared 46MB L3 cache, and 128GB main memory. 
\tool{} is compatible with JDK 1.5 and any of its successors. We run all applications with JDK 1.8.0\_161.

\paragraph{Applications and Benchmarks.} The lightweight nature of \tool{} allows us to collect profiles from a variety of Java and Scala applications obtained from the Awesome Java repository~\cite{awesome-java}, such as the Renaissance benchmark suite~\cite{Prokopec:2019:RBS:3314221.3314637}, Soot~\cite{soot}, parquet MR~\cite{parquet-mr}, Findbugs~\cite{FindBugs}, Eclipse Deeplearning4J~\cite{deeplearning4j}, JGFSerialBench~\cite{grande}, RoaringBitmap~\cite{RoaringBitmap}, Apache SAMOA~\cite{SAMOA}, to name a few.
We run these applications with different real inputs released with them or the real inputs that we can find to our best knowledge; the inputs control the parallelism configuration.

\begin{table*}
\centering
\includegraphics[width=\textwidth]{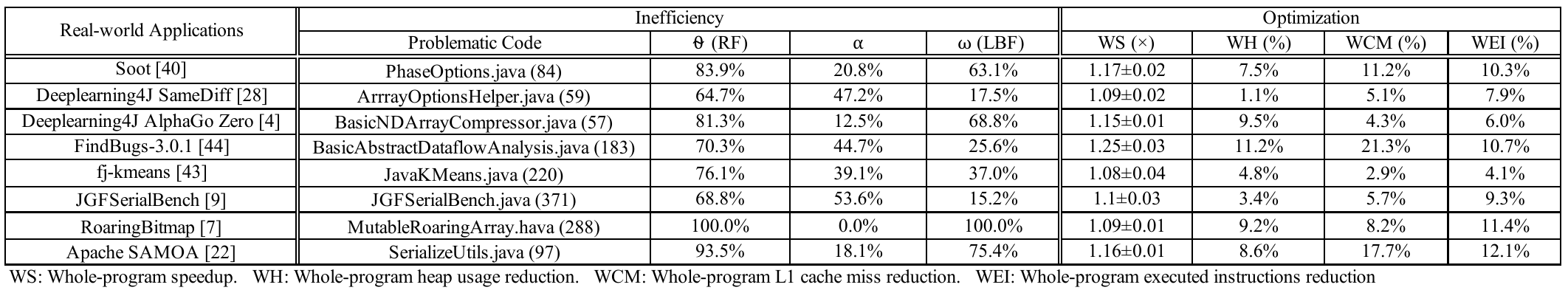}
\caption{Overview of performance optimization guided by \tool{}.}
\label{overview}
\end{table*}

\paragraph{Replication.}

Figure~\ref{redundancy ratio} shows the replication ratios for more than 50 Java programs obtained from \tool{}. We can see that several Java programs suffer from significant object replications (replication ratio > 15\%). We optimize some of them, as shown in Section~\ref{case study} under the guidance of \tool{}. For some Java applications (e.g., gauss-mix, log-regression, page-rank, scala-kmeans) with high replication ratios, we only obtain trivial speedups because these applications do not have any hotspot object replicas. For example, the top five object replicas' accessing times in Renaissance benchmark gauss-mix are less than three. In this case, it is reasonable that there is no benefit to optimize these replicated objects that are used very few. We focus on the hotspot object replicas, which at least are accessed dozens of times.

\tool{} is able to pinpoint many object replicas that are not reported by existing profilers and guide optimization choices. 
Table~\ref{overview} summarizes the new findings identified by \tool{}, which we further elaborate in Section~\ref{case study}.
In Table~\ref{overview}, we report replication factor $\theta$, $\alpha$, and lower bound factor $\omega$, which are defined in Section~\ref{theory}. 
Table~\ref{overview} shows that $\alpha$ ranges from $0\%$ to $53\%$ and $\theta$ ranges from $65\%$ to $100\%$, respectively. 
As a result, the lower bound factor $\omega$ is usually $>15\%$, which means that these Java applications at least have $15\%$ objects suffering from object replication. 

\paragraph{Optimization.}
It is worth doing the optimization to decrease the creations of objects with  the same contents.
To guarantee optimization correctness, we ensure the optimized codes do not change semantics for any inputs and pass the validation tests.
To avoid system noises, we run each application 30 times and use a 95\% confidence interval for the geometric mean speedup to report the performance improvement, according to a prior approach~\cite{inefficiencies-in-java}. 

From Table~\ref{overview}, we can see that we are able to obtain nontrivial speedups by removing object replicas. The performance improvement comes from the reduction of heap memory usage, cache misses, and executed instructions, which are measured with jmap~\cite{jmap} and perf~\cite{perf}. We detected the object replicas as shown in Table~\ref{overview} without much effort. Object replicas often concentrate around only a few calling contexts making investigation relatively simpler; for example, in all of our case studies, we found the top five objects (sorted by replication factor) account for $\sim$37\% of whole-program object replicas on average.

\begin{figure*}[t]
\begin{center}
\begin{subfigure}[b]{0.499\textwidth}
\includegraphics[width=\textwidth]{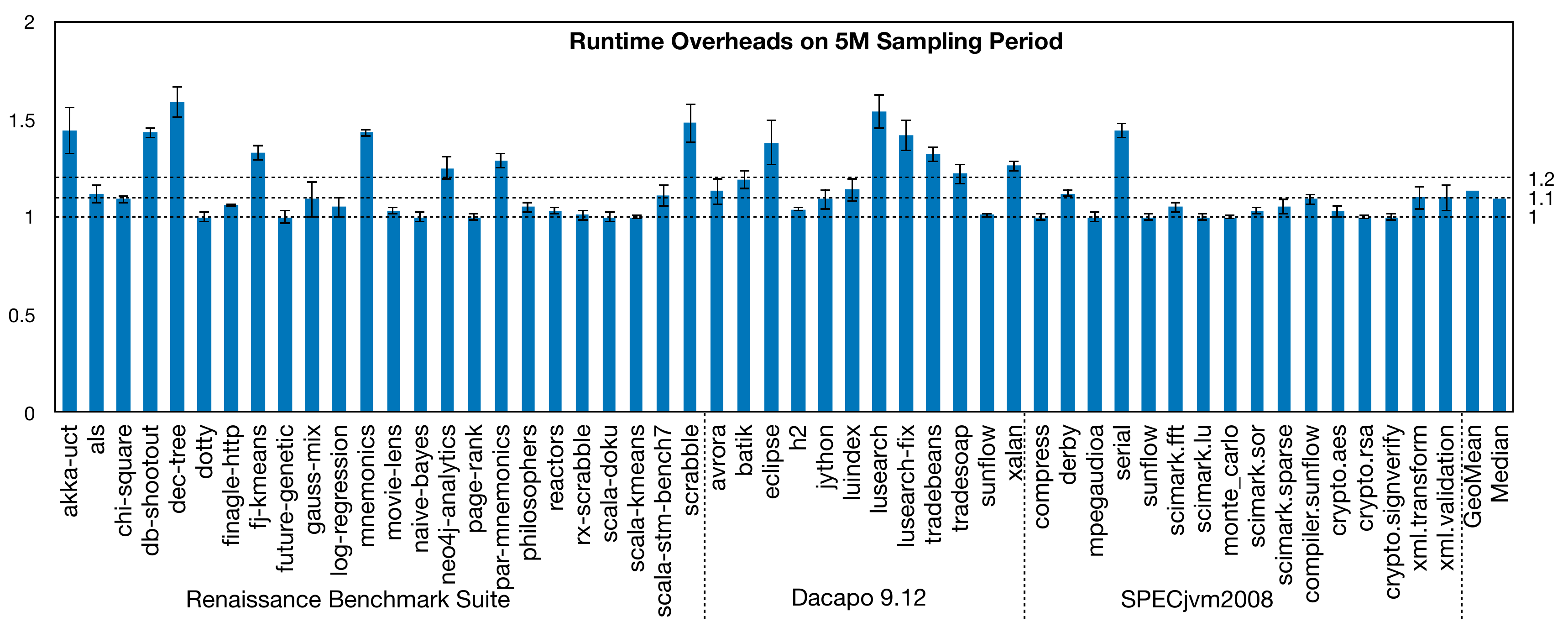}
\caption{Runtime overheads.}
\label{fig:runtime-slowdown}
\end{subfigure}
~
\begin{subfigure}[b]{0.499\textwidth}
\includegraphics[width=\textwidth]{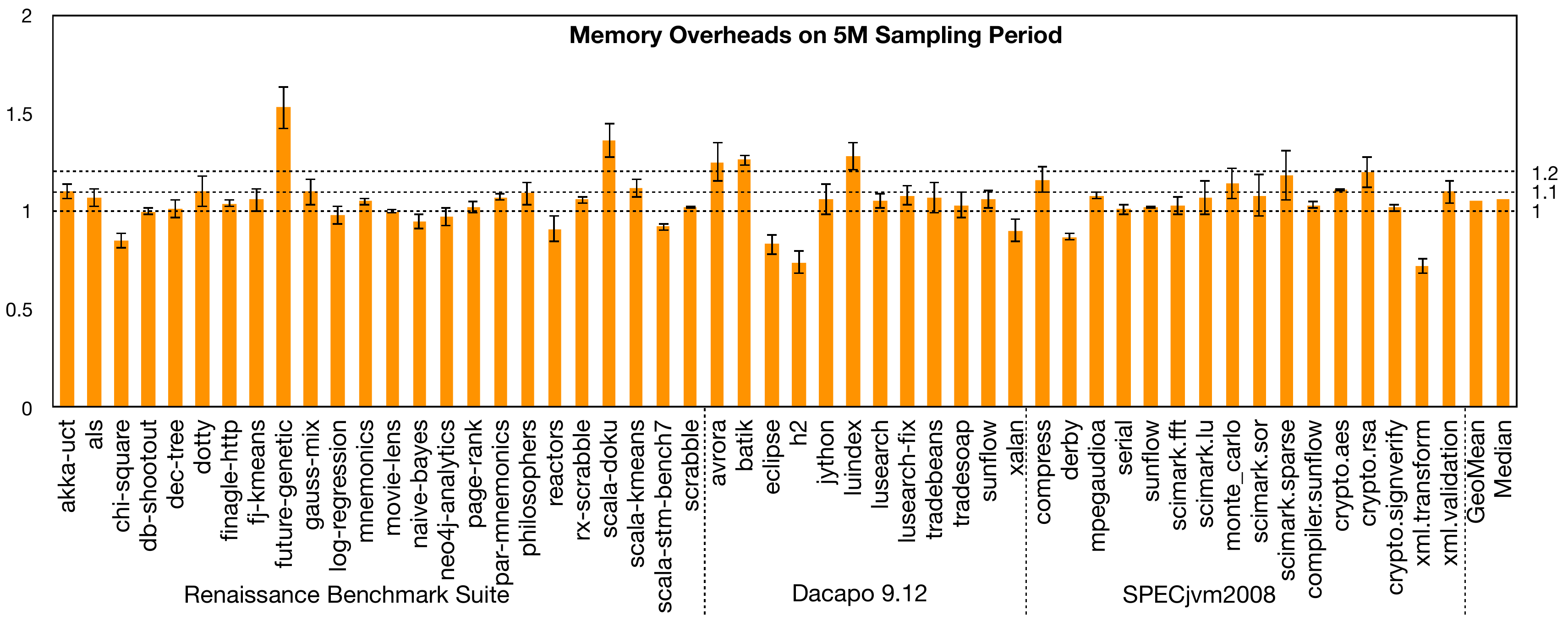}
\caption{Memory overheads.}
\label{fig:memory-bloat}
\end{subfigure}
\end{center}
\vspace{-0.1in}
\caption{\tool{}'s runtime and memory overheads in the unit of times ($\times$) on various benchmarks.}
\label{fig:overhead-5M}
\end{figure*}

\subsection{False Positives and Negatives}
\label{false positive}
As a sampling-based tool, \tool{} can introduce false negatives --- missing some object replicas. However, the statistics theory guarantees the high probability of capturing object replicas that occur frequently. The false negatives do not hurt the insights obtained from \tool{} because optimizing infrequently occurred object replicas typically receives trivial speedups.

\tool{} can also introduce false positives --- reporting object replicas that are not replicated because \tool{} uses sampling instead of exhaustive checking.  
The false positives occur when most elements of the two objects are the same, with only a few different elements not monitored.
However, \tool{} randomly checks different fields with enough samples to minimize the false positives. 
We exhaustively check every field of the top five objects (i.e., objects associated with most samples) of all our investigated programs in Table~\ref{false-positive}. From the table, we can see that \tool{} incurs 5.9\% false positives. 



\begin{table}[t]
\small
\centering
\label{false-positive}
\begin{tabular}{|l|l|c|c|}
\hline
\multicolumn{2}{|l|}{\multirow{2}{*}{\textbf{}}} & \multicolumn{2}{c|}{\textbf{Replica Detection}} \\ \cline{3-4}
\multicolumn{2}{|l|}{} & \textbf{Replicated} & \textbf{Not Replicated} \\ \hline
\multirow{2}{*}{\textbf{Actual}} & \textbf{Replicated} & 8 & 0 \\ \cline{2-4}
 & \textbf{Not Replicated} & 3 & 48 \\ \hline
\hline
\multicolumn{2}{|l|}{Correctness} & \multicolumn{2}{l|}{(48 + 8) / (0 + 8 + 3 + 48) = 94.9\%} \\ \hline
\multicolumn{2}{|l|}{False positive rate} & \multicolumn{2}{l|}{3 / (3 + 48) = 5.9\%} \\ \hline
\end{tabular}
\caption{Accuracy for \tool{}'s replica detection.}
\label{false-positive}
\vspace{-1em}
\end{table}

\subsection{Overhead Measurement}
The runtime overhead (memory overhead) is the ratio of the runtime (peak memory usage) of the execution monitored by \tool{} to the runtime (peak memory usage) of the native execution. To quantify the overhead, we apply \tool{} to three well-known Java benchmark suites: Renaissance~\cite{Prokopec:2019:RBS:3314221.3314637}, Dacapo 9.12~\cite{dacapo-9.12}, and SPECjvm2008~\cite{specjvm2008}. We run all benchmarks with four threads.
We run every benchmark 30 times and compute the average and error bar.
Figure~\ref{fig:overhead-5M} shows the overhead when \tool{} is enabled at a sampling period of 5M. Some Renaissance and Dacapo benchmarks have higher time overhead (larger than 30\%) because they allocate too many objects (e.g., more than $400$ million allocations for mnemonics, par-mnemonics, scrabble, akka-uct, db-shootout, dec-tree, neo4j-analytics). 

\section{Case Studies}
\label{case study}

This section shows how \tool{} pinpoints object replicas in real applications and guides the optimization. Our optimization guarantees the program's correctness via human inspection, and we have evaluated our transformed code with tests to ensure their correctness.
It is worth noting that existing profilers may identify the same object allocation as a hotspot in memory usage; however, they do not know whether this allocation point creates multiple object replicas for potential optimization. \tool{}, in contrast, quantifies the replication factors for the objects to provide intuitive optimization.
We have submitted our optimization patches in several cases and gotten them confirmed or upstreamed, e.g., Soot and Findbugs. 


\subsection{Soot}
\begin{figure}
\centering
\includegraphics[width=.49\textwidth]{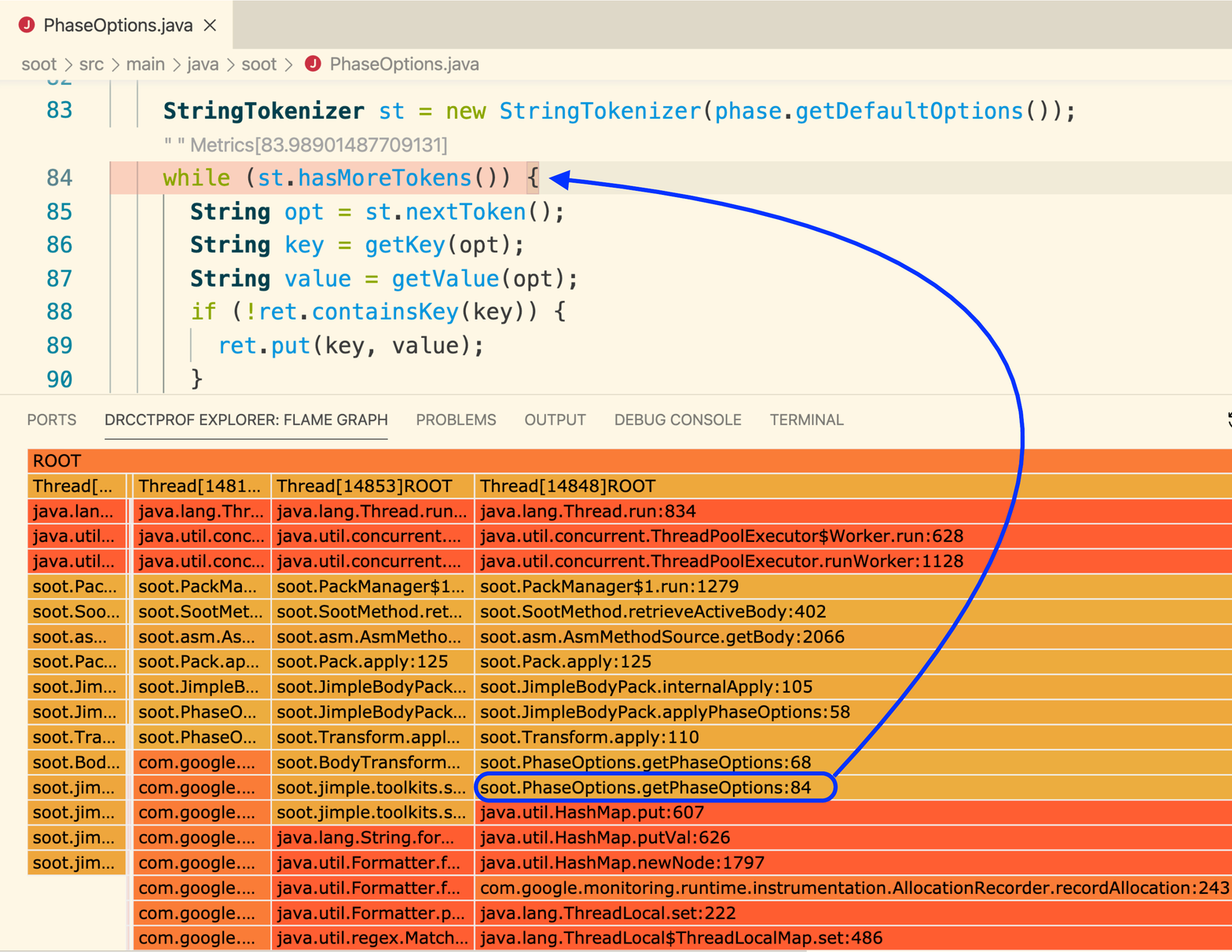}
\caption{The Flame Graphs GUI of Soot shows a problematic object {\tt st} with its accesses in full call stacks.}
\label{gui}
\vspace{-1em}
\end{figure}

Soot is a Java optimization framework, which uses containers extensively~\cite{soot}. 
We run Soot-3.3.0 using the bytecode of the DaCapo benchmark avrora as input. 
Figure~\ref{gui} shows the snapshot of \tool{}'s Flame Graphs GUI in VSCode for intuitive analysis. The top pane of the GUI shows the Java source code; the bottom shows the flame graphs of object accesses in their full call stacks. In the flame graphs, the x-axis shows the accesses with their call stacks to object replicas, and the y-axis shows call stack depth, counting from zero at the top. Each rectangle represents a stack frame. The wider a stack frame is, the higher of replication factor of this stack frame. The GUI in Figure~\ref{gui} shows one problematic object {\tt st} (highlighted in blue), which is accessed on line 84 in method {\tt getPhaseOptions} of class {\tt PhaseOptions} with many replicas (its replication factor $\theta$ is 83.9\%, as shown in the top pane of the GUI). 

Soot's execution is divided into a number of phases, such as Jimple Body Creation (jb) phase, Java To Jimple Body Creation (jj) phase, Grimp Body Creation (gb) phase, etc. In the jb phase, the JimpleBodys are built by a phase called jb, which is itself comprised of subphases, such as the aggregation of local variables (jb.a), type assigner (jb.tr), dead assignment eliminator (jb.dae), etc. Each of these subphases that belong to jb phase has its own default option. By investigating the source code, we found that when the soot executes these subphases sequentially, the default option for different subphases is stored in the reported StringTokenizer {\tt st} object, which keeps unchanged. 

To eliminate these replicas, we only read the default option when its contents are changed; otherwise, we reuse the default option from the prior subphase. This optimization yields a $(1.17\pm 0.02)\times$ speedup to the entire program.

\subsection{Eclipse Deeplearning4J -- SameDiff}
Eclipse Deeplearning4J integrates with Hadoop and runs on several backends~\cite{deeplearning4j}. We run Deeplearning4J using SameDiff, a TensorFlo-\\w/PyTorch-like framework for executing complex graphs. This framework is also the lower lever base API for running onnx and TensorFlow graphs.
\tool{} investigates the training phase and reports that the replication factor $\theta$ of the input array {\tt shapeInfo} is 64.7\%, indicating high redundancies in the computation on this array in method {\tt hasBitSet}, which determines array types, as shown in Listing~\ref{deeplearning listing}.  

Deeplearning4J SameDiff builds a directed acyclic graph, whose nodes are differential functions used to compute gradients. In the SameDiffLayer (a base layer used for implementing Deeplearning4J layers with SameDiff), Deeplearning4J provides a set of operations named "Custom operations" designed for the SameDiff graph. To execute the "Custom operations" within graph, these operations are stored in a two-dimensional array. Then Deeplearning4J splits this two-dimensional array into different small partitions. Each partition has its own shape identifier array {\tt shapeInfo}, which is used to determine four shape properties: {\tt SPARSE}, {\tt COMPRESSED}, {\tt EMPTY}, and {\tt DENSE}. We found that the adjacent partitions in the SameDiff graph often have the same shape property due to the good locality among adjacent partitions.

To eliminate redundancies, we first check whether the {\tt shapeInfo} in the current iteration has the same value as in the last iteration. If the {\tt shapeInfo} is unchanged, we reuse the shape property memoized from the previous iteration, which saves the call to {\tt hasBitSet}.
This  yields a $(1.09\pm0.02)\times$ speedup to the entire program.

\begin{figure}
\begin{lstlisting}[firstnumber=58,language=java]
public static ArrayType arrayType(long[] shapeInfo) {
  @$\blacktriangleright$@val opt = Shape.options(shapeInfo);
  if (hasBitSet(opt, ATYPE_SPARSE_BIT))
    return ArrayType.SPARSE;
  else if (hasBitSet(opt, ATYPE_COMPRESSED_BIT))
    return ArrayType.COMPRESSED;
  else if (hasBitSet(opt, ATYPE_EMPTY_BIT))
    return ArrayType.EMPTY;
  else
    return ArrayType.DENSE;
}
\end{lstlisting}
\vspace{-0.3in}
\captionof{lstlisting}{\tool{} identified the {\tt shapeInfo} array with replicas in Deeplearning4J SameDiff.}
\vspace{-0.1in}
\label{deeplearning listing}
\end{figure}

\subsection{Eclipse Deeplearning4J -- AlphaGo Zero}
We also run Deeplearning4J using AlphaGo Zero model~\cite{zero}, which combines a neural network and Monte Carlo Tree Search in an elegant policy iteration framework to achieve stable reinforcement learning. \tool{} studies the training stage and reports an object, {\tt Map<String, NDArrayCompressor> codecs}, which is allocated on line 53 and accessed on line 57 in method {\tt loadCompressors} of class {\tt BasicNDArrayCompressor} with many replicas (its replication factor $\theta$ is 81.3\%), as shown in Listing~\ref{zero}.

The reason for generated replicas is due to constructing the computational graph in the AlphaGo Zero model. To initialize the computational graph, the AlphaGo Zero model uses an existing array {\tt parameters} for each layer. Given the topological order, the AlphaGo Zero model constructs the computational graph by iterating each subset of array {\tt parameters}. Since the array {\tt parameters} is in compressed status, to obtain the elements of array {\tt parameters}, the program needs to decompress it first. Deeplearning4J framework provides several different compression algorithms (compressor) based on the data type (e.g., {\tt FLOAT16}, {\tt FLOAT8}, {\tt INT16}, etc). Since every subset of array {\tt parameters} has the same data type, which means we don't need to load the new compressor and store it again in map {\tt codecs} in a loop (line 69 of Listing~\ref{zero}). To avoid the redundant loading and storing compressor, we check whether the program is processing  different subsets in the same array {\tt parameters}. If so, we use the current compressor directly. This optimization yields a $(1.15\pm0.01)\times$ speedup to the entire program.

\begin{figure}
\begin{lstlisting}[firstnumber=58,language=java]
public void init(INDArray parameters) {
  ...
  for (int vertexIdx : topologicalOrder) {
      paramsViewForVertex[vertexIdx] = parameters.get(NDArrayIndex.interval(0,0,true));
      //get method calls loadCompressors to decompress data
  }
}
@$\blacktriangleright$@protected Map<String, NDArrayCompressor> codecs = new ConcurrentHashMap<>();
protected void loadCompressors() {
  ...
  for (NDArrayCompressor compressor : compressors) {
    @$\blacktriangleright$@codecs.put(compressor.getDescriptor(), compressor);
  }
}
\end{lstlisting}
\vspace{-0.3in}
\captionof{lstlisting}{\tool{} identified the {\tt codecs} map with replicas in Deeplearning4J AlphaGo Zero.}
\vspace{-0.1in}
\label{zero}
\end{figure}

\subsection{FindBugs-3.0.1}
FindBugs looks for code instances that are likely to be errors~\cite{FindBugs}. We run Find-Bugs on a real input Java chart library 1.0.19 (a widely used client-side chart library for Java). \tool{} reports an object that has many replicas, {\tt BasicBlock block} (an object with a user-defined type), which is accessed on the line 183 in method {\tt lookupOrCreateFact} of class {\tt BasicAbstractDataflowAnalysis}, as shown in Listing~\ref{findbug listing}.

The replicas come from the algorithm of data-flow analysis used in FindBugs. Findbugs divides a data-flow graph into tiny-sized blocks and creates an object for each block instead of creating a single object for the whole graph. Consequently, most created objects have the same content due to good value locality among adjacent blocks. \tool{} finds that the method {\tt lookupOrCreateFact} (line 182 of Listing~\ref{findbug listing}) method is usually invoked with the same input {\tt BaiscBlock block}. \tool{} reports that the replication factor $\theta$ of the input {\tt block} is 70.3\%, indicating many replicas of this object. To avoid the redundant lookup and creation, we check whether a different {\tt block} is produced in the current iteration. If the {\tt block} is unchanged, we return {\tt fact} obtained from the last invocation directly. This optimization yields a $(1.25\pm0.03)\times$ speedup to the entire program.

\begin{figure}
\begin{lstlisting}[firstnumber=176,language=java]
public/* final */Fact getStartFact(BasicBlock block) {
  return lookupOrCreateFact(startFactMap, block);
}
public/* final */Fact getResultFact(BasicBlock block) {
  return lookupOrCreateFact(resultFactMap, block);
}
private Fact lookupOrCreateFact(Map<BasicBlock, Fact> map, BasicBlock block) {
  @$\blacktriangleright$@Fact fact = map.get(block);
  if (fact == null) {
    fact = createFact();
    map.put(block, fact);
  }
  return fact;
}
\end{lstlisting}
\vspace{-0.3in}
\captionof{lstlisting}{The source code highlighed by \tool{} shows the object {\tt block} with replicas in Findbugs.}
\vspace{-0.1in}
\label{findbug listing}
\end{figure}

\subsection{fj-kmeans}
fj-kmeans is a benchmark from Renaissance Suite, used to run the k-means algorithm~\cite{Prokopec:2019:RBS:3314221.3314637}. We run fj-kmeans using the fork/join framework as input. \tool{} reports an object that has many replicas, array {\tt result}, which is allocated on line 5 in method {\tt findNearestCentroid} and accessed on line 2 in method {\tt compute\\Directly} of class {\tt JavaKMeans}, as shown in Listing~\ref{fj-kmeans listing}.

The generated replicas are due to the finding nearest centroid algorithm for many different sets of elements. This finding nearest centroid algorithm maintains a collection of centroids, and the distance between each centroid is significant. Then, during some computation periods, because different sets of elements have small distance, the program keeps generating the same centroids and put the centroids' indices into an array {\tt result} (line 12 of Listing~\ref{fj-kmeans listing}), which is the object with many replicas (the replication factor $\theta$ is 76.1\%) reported by \tool{}. 

To eliminate efficiencies, we check the values in array {\tt result} produced by {\tt findNearestCentroid}. If {\tt result} is unchanged, we reuse the return value of method {\tt collectClusters} memoized from the last iteration, which avoids the redundant computation. This optimization yields a $(1.08\pm0.04)\times$ speedup to the entire program.

\begin{figure}
\begin{lstlisting}[firstnumber=1,language=java]
protected Map<Double[], List<Double[]>> computeDirectly() {
  @$\blacktriangleright$@return collectClusters(findNearestCentroid());
}
private int[] findNearestCentroid() {
  @$\blacktriangleright$@final int[] result = new int[taskSize];
  ...
  for(...) {
  final Double[] element = data.get(dataIndex);
    for(...) {
      final double distance = distance(element, centroids.get(centroidIndex));
      ...
      result[dataIndex - fromInclusive] = centroidIndex;
    }
  }
  return result;
}
private Map<Double[], List<Double[]>> collectClusters(final int[] centroidIndices) {
  //computation with input parameter centroidIndices
}
\end{lstlisting}
\vspace{-0.3in}
\captionof{lstlisting}{\tool{} pinpoints the {\tt result} object with many replicas in fj-kmeans. \vspace{-1em}}
\label{fj-kmeans listing}
\end{figure}

\section{Threats to Validity}
\label{validity}

The threats reside in validating \tool{}'s optimization guidance.
The limited scope of replica detection and the sampling strategy does not reveal the ground truth of object replication.
Any reported replication is input and execution specific.
Different inputs can result in different profiles, and the effects of optimization on unseen inputs will remain unknown.
In our studies, we use real inputs for the applications to ensure the optimization is valid. 
Finally, our optimization may sometimes break the program readability  by inserting conditional checks. Developers need to decide whether to adopt our optimization given their priority on software performance.

\section{Conclusions}
\label{conclusion}

In this paper, we design and develop \tool{}, the first lightweight profiler to identify object replicas in Java applications.  
As a unique feature, \tool{} combines the use of performance monitoring units, debug registers and lightweight byte code instrumentation for statistical object replica detection.  
With the evaluation of more than 50 Java applications, we show \tool{} minimizes false positives and incurs 9\% and 6\% runtime and memory overheads, respectively. 
We further optimize several real-world applications guided by \tool{} that result in a noticeable reduction in heap-memory demands and significant runtime speedups.
Many optimization patches are confirmed or upstreamed by the software developers. \tool{} is open source at \url{https://github.com/Xuhpclab/jxperf}.


\section*{Acknowledgements}
We thank the anonymous reviewers for their valuable comments. This research was supported by NSF 2050007 and a Google gift.

\bibliography{datacentric}
\bibliographystyle{ACM-Reference-Format}

\end{document}